# Solving nonlinear PDEs with Quantum Neural Networks: A variational approach to the Bratu Equation

**Nikolaos Cheimarios[1]**

School of Chemical Engineering, National Technical University of Athens, Athens 15780, Greece

e-mail: nixeimar@chemeng.ntua.gr

**Abstract**

We present a variational quantum algorithm (VQA) to solve the nonlinear one-dimensional Bratu equation. By formulating the boundary value problem within a variational framework and encoding the solution in a parameterized quantum neural network (QNN), the problem reduces to an optimization task over quantum circuit parameters. The trial solution incorporates a predictor from the previous continuation step and boundary-enforcing terms, allowing the circuit to focus on minimizing the residual of the differential operator. Using a noiseless quantum simulator, we demonstrate that the method accurately captures both solution branches of the Bratu equation and shows excellent agreement with classical pseudo arc-length continuation results.

## 1. Introduction

The rapid development of quantum computing has opened transformative possibilities for solving complex computational problems that challenge classical hardware. Among the most promising directions is the application of quantum algorithms to the numerical solution of partial differential equations, particularly those arising in physics and engineering. Solving partial differential equations (PDEs) is a cornerstone of scientific computing, with applications spanning fluid dynamics, quantum mechanics, and financial modeling. In recent years, the potential of quantum computing for tackling PDEs has garnered significant attention due to its promise of exponential speedups for certain linear algebra subroutines [1–5]. Beyond linear systems,

---

[1] Current Address: Accenture, Arcadias Str., 1, Kifissia, Athens, Greece, 14564

approaches such as variational quantum eigensolvers (VQE) [6,7] and quantum annealing [8] have been explored for nonlinear PDEs [9–11] and optimization-based formulations [12], respectively. While these methods hold theoretical appeal, practical challenges persist. These include the overhead of state preparation [13–15], noise in current quantum hardware [16–19], and the difficulty of encoding nonlinear terms into quantum frameworks [20]. Recent work has shown that nonlinear dynamics can be embedded into variational quantum frameworks, either by formulating nonlinear problems as variational optimization tasks or by mapping their solution structure onto quantum ground states, highlighting the relevance of variational quantum algorithms for nonlinear scientific computing [21,22]. While current quantum processors face limitations in terms of qubit count, noise, and error correction, they are reaching a level of maturity that allows meaningful experiments on small-scale, structured problems. Algorithms that are variational in nature—especially those based on VQE — offer a viable path for practical quantum computation in the near term.

In this context, quantum-ready implementations—algorithms that are directly compatible with real quantum hardware—are crucial. These approaches are designed to work within the constraints of current or near-future devices, while exploiting their ability to optimize complex functions via parameterized quantum circuits. This work investigates the application of a variational quantum algorithm to solve the Bratu equation, a nonlinear boundary value problem known for exhibiting bifurcation behavior. Due to its mathematical structure and physical relevance in phenomena like thermal ignition [23], the Bratu equation serves as a standard benchmark for evaluating solution techniques.

We formulate a quantum-ready solution method using a quantum neural network (QNN) as the trial function in a variational framework. The circuit design, parameter initialization, and optimization process are aligned with the requirements of near-term quantum hardware. Simulations carried out using an ideal quantum backend demonstrate the method's accuracy and ability to resolve both branches of the Bratu solution. These results support the growing feasibility of quantum PDE solvers and provide a concrete step toward their use in nonlinear scientific computing.

## 2. Problem formulation

In this section, we present the mathematical formulation of the nonlinear Bratu equation and describe the variational quantum approach employed to approximate its solution. The core idea is to represent the solution using a parameterized quantum model embedded within a trial function that inherently satisfies the boundary conditions. This formulation enables the conversion of the differential equation into an optimization problem over quantum circuit parameters. By minimizing the residual of the differential operator, we seek to find a quantum-enhanced approximation of the true solution. The following subsections detail the structure of the Bratu equation, the design of the quantum trial solution, and the optimization procedure used to train the underlying QNN.

### 2.1 The Bratu equation

The 1D Bratu equation is a nonlinear boundary value problem defined as,

$$\frac{d^2u}{dx^2} + \lambda e^u = 0, x \in (0,1) \tag{1}$$

with boundary conditions $u(0) = u(1) = 0$, where $u(x)$ is the unknown solution, $\lambda$ is a parameter controlling the strength of the exponential term.

### 2.1 Quantum Trial Solution

The core of our variational quantum approach lies in the computation of the quantum model output $u_q(x; \theta)$, which plays a central role in constructing a trial solution to the Bratu equation. $u_q(x; \theta)$ is realized as a QNN - a parameterized quantum circuit whose structure and training process are inspired by classical neural networks, but which leverages the principles of quantum mechanics to achieve enhanced expressivity and potential computational advantages.

We define the trial function of the Bratu equation as,

$$u_{trial}(x; \theta) = u_{pred}(x) + s \cdot x(1-x) \cdot u_q(x; \theta) \tag{2}$$

where $s$ is a scaling factor introduced to control the amplitude. The term $u_{pred}(x)$ currently denoted is a predictor taken from the previous continuation step; after initialization, it is always the previously computed quantum solution. The upper branch of the Bratu problem corresponds to an unstable solution, for which residual-based optimization exhibits strong convergence basin sensitivity; the inclusion of a predictor term is therefore essential to steer the variational optimization toward this solution. Eq. (2) without the term $u_{pred}(x)$ is sufficient for capturing the entire lower branch. However, to direct the optimizer towards the upper solution branch $u_{pred}(x)$ is necessary and is conveniently given by the solution of the previous step. The prefactor $x(1-x)$ ensures that the trial solution satisfies the boundary conditions $u(0) = u(1) = 0$ automatically, regardless of the quantum circuit output. This approach reduces the burden on the quantum model by eliminating the need to learn boundary behavior explicitly and focuses optimization on minimizing the residual of the governing differential equation in the interior of the domain. The role of the quantum circuit is thus to provide a flexible, parameterized function $u_q(x; \theta)$ whose form can be tuned during optimization to minimize the residual of the Bratu differential equation.

The computation of $u_q(x; \theta)$, proceeds through several stages, analogous to the layers of a classical feedforward neural network. The scalar input $x \epsilon [0,1]$ is first transformed into a feature space using a nonlinear encoding function,

$$\varphi(x) = [\sin(\pi x),\ e^{-10(x-0.5)^2},\ \sin(3\pi x)] \tag{3}$$

The choice of feature functions is critical for guiding the quantum circuit to learn meaningful corrections to the classical Bratu solution. The term $\sin(\pi x)$ corresponds to the fundamental sine mode which captures the lowest frequency spatial variation. This mode often appears in analytical solutions of boundary value problems and provides a smooth global feature across the domain. To complement this, we include the localized function $e^{-10(x-0.5)^2}$, a narrow Gaussian centered at $x = 0.5$. This term enables the ansatz to represent sharp peaks or localized features in the solution — a critical capability when capturing upper-branch Bratu solutions, which typically exhibit steep gradients or blow-up behavior near the center of the domain. Finally, the inclusion of $\sin(3\pi x)$ represents high curvature near center and introduces higher-frequency content, allowing the model to express finer-scale oscillatory corrections. Together, these encoding terms provide a compact yet expressive representation of spatial structure, enabling the variational quantum model to correct the classical solution efficiently without overparameterization. This three-dimensional feature vector designates the symmetry parabolic shape of the solution of the 1D-Bratu equation and serves as the input layer of the QNN.

Next, this feature vector is used to initialize a quantum state across three qubits. Each qubit undergoes a rotation around the Y-axis, with the rotation angle taken cyclically from the feature vector. That is, for qubit $i$, the operation $R_Y(\theta_i)$ is applied, where $\theta_i = \varphi(x)[i \bmod 3]$. This stage embeds the classical input into the quantum Hilbert space by preparing a parameterized superposition state $|\psi_x\rangle$.

Following this input encoding, the quantum circuit applies a sequence of trainable unitary operations, structured as Strongly Entangling Layers in the context of PennyLane [24], which together constitute the hidden layers of the QNN. Each layer consists of local single-qubit rotations around the X, Y, and Z axes, parameterized by learnable weights $\theta$, followed by a fixed pattern of entangling gates (typically controlled-NOT operations) that generate quantum correlations among the qubits. If the circuit contains $L$ such layers and $n$ qubits, the trainable parameters are represented by a three-dimensional tensor $\theta \in \mathbb{R}^{L \times n \times 3}$, analogous to the weights in a deep neural network.

These operations define a full unitary transformation $U(x; \theta)$ that acts on the initial quantum state, resulting in a final parameterized state,

$$|\psi_\theta(x)\rangle = U(x; \theta)|0\rangle^{\otimes n} \tag{4}$$

The output of the quantum circuit is then defined as the expectation value of The Pauli-Z operator, denoted, on the last qubit, $(n - 1)$, in the system,

$$u_q(x; \theta) = \langle \psi_\theta(x)|Z_{n-1}|\psi_\theta(x)\rangle \tag{5}$$

This produces a scalar value in the range $[-1, 1]$, which is used as the nonlinear modeling component in the trial solution. Alternatively, this quantity can be written as,

$$u_q(x;\theta) = \langle 0|^{\otimes n} U^\dagger(x;\theta) Z_{n-1} U(x;\theta)|0\rangle^{\otimes n} \tag{6}$$

emphasizing the action of the unitary evolution and measurement on the initialized quantum state.

The quantum circuit operates as a QNN: the input $x$ is transformed into a quantum state, passed through a series of trainable unitary transformations (analogous to layers in a neural network), and then measured to produce a classical scalar output. The trainable parameters $\theta$ are optimized using a classical gradient-based optimizer to minimize a cost function derived from the residual of the Bratu differential equation.

## 2.2 Variational Optimization of the Quantum Neural Network

The optimization objective is to minimize a cost function based on the integrated squared residual of the Bratu equation over the domain [0,1]. Since the differential equation involves the second derivative of $u(x)$, we approximate $\frac{d^2 u}{dx^2}$ numerically using a central finite-difference formula,

$$\frac{d^2 u(x;\theta)}{dx^2} = \frac{u_{i+1} - 2u_i + u_{i-1}}{h^2} \tag{7}$$

where $h$ is the discretization step.

Using this approximation, we define the residual function,

$$R_\theta(x) = \frac{d^2 u(x;\theta)}{dx^2} + \lambda e^{u(x;\theta)} \tag{8}$$

and the corresponding cost function is the $L^2 - norm$ of the residual,

$$C_\theta(x) = \int_0^1 [R_\theta(x)]^2 dx \tag{9}$$

The parameter optimization is carried out using a classical gradient-based optimizer. In our implementation, we employ the Adam optimizer, a stochastic gradient descent method with momentum and adaptive learning rates. Compared to other optimizers available, this optimizer is particularly well suited to the noisy and non-convex optimization landscapes that arise in variational quantum algorithms. At each iteration, the parameters $\theta$ are updated according to,

$$\theta^{k+1} = \theta^k - \eta \cdot \hat{g}^k \tag{10}$$

where $\eta$ is the learning rate and $\hat{g}^{(k)}$ is the bias-corrected gradient estimate computed from the cost function $C_\theta(x)$ at iteration $k$. The gradients of the cost function with respect to the quantum circuit parameters are computed using automatic differentiation, leveraging the parameter-shift rule available in the quantum software framework.

The optimization process is initialized with a set of variational parameters $\theta^{(0)}$ chosen to bias the trial solution toward a particular solution branch. For instance, to favor convergence to the upper (large amplitude) solution of the Bratu equation, we initialize the rotational angles in the circuit with relatively high values (e.g., near 2.0 radians), thereby providing a strong initial guess in the relevant solution basin.

The optimization loop proceeds for a fixed number of iterations or until convergence is observed in the cost function. Throughout training, the history of the cost value $C_\theta(\theta^k)$ is recorded to monitor progress and assess stability. Upon convergence, the optimized parameters $\theta^*$ define the final QNN circuit, from which the approximate solution $u_q(x; \theta^*)$ is computed and evaluated.

## 3. Computational details

All simulations are performed using the PennyLane quantum machine learning framework, which facilitates hybrid quantum-classical computations. All evaluations of the variational trial function are obtained directly from measurements of the parameterized quantum circuit executed on the quantum simulator; no classical surrogate model is used to evaluate the quantum circuit output. The quantum circuits is executed on the `default.qubit` simulator backend provided by PennyLane, which emulates an ideal, noise-free quantum processor. A total of *n*=3 qubits are used to represent the quantum state and fit the encoding functions (see Eq. 3). In the context of the encoding functions used within the circuit, the selected number represents an optimal balance between representational power and computational efficiency. Empirically, this particular number of encoding functions is sufficient to capture the essential structure of the solution, including its most challenging features such as the sharp gradients and high-curvature behavior near the turning point. Increasing the number of encoding functions beyond this point results in only marginal improvements in accuracy while substantially increasing the computational burden, both in terms of training time and resource usage. On the other hand, reducing the number of encoding functions leads to underfitting, with the model failing to converge—especially near critical regions like the turning point where solution behavior changes rapidly and requires higher expressivity. This sensitivity highlights the role of the encoding functions in enabling the circuit to distribute its representational capacity adaptively across the domain. The selected configuration emerges as a minimal yet sufficient encoding, tuned to the complexity of the problem: expressive enough to achieve convergence and precise approximation of the upper solution, yet compact enough to avoid unnecessary overhead. Thus, this choice is not arbitrary but reflects a principled tradeoff between convergence, accuracy, and efficiency.

The parameterized quantum model is composed of $L=4$ layers of Strongly Entangling Layers. Each layer applies trainable single-qubit rotations around the X, Y, and Z axes, followed by a fixed entangling pattern of controlled-NOT (CNOT) gates. The total number of trainable parameters in the quantum neural network is thus $3\times4\times3=36$. The number of layers in the circuit is carefully chosen to strike a crucial balance between depth-driven expressivity and computational tractability. A shallow architecture lacks the hierarchical capacity needed to approximate the intricate structure of the upper solution, particularly in regions of high curvature or near the turning point, where subtle but significant variations demand deeper representations. Empirical observations confirm that circuits with fewer layers fail to converge or yield approximations with significant errors in these critical regions. Conversely, increasing the number of layers beyond the chosen value does not significantly enhance accuracy but introduces greater computational cost and a risk of overfitting or optimization instability. Deeper architectures also tend to suffer from vanishing gradients and increased training time, which outweigh the minor gains in performance. The selected depth is therefore neither arbitrary nor heuristic—it reflects an informed choice based on the observed dynamics of convergence and expressivity. It is deep enough to allow the circuit to model complex, nonlinear features of the solution, yet shallow enough to remain efficient and stable during training. This choice enables the circuit to faithfully represent the target behavior with minimal architectural overhead, contributing to both the robustness and efficiency of the overall method.

The input to the quantum circuit consists of a one-dimensional scalar $x \in [0,1]$, which is first transformed into a three-dimensional nonlinear feature vector. This feature vector is then used to initialize the quantum state through $R_Y$ rotations. The trial solution $u_{trial}(x;\theta)$ is constructed to satisfy the boundary conditions by incorporating the factor $x(1-x)$, while the quantum output $u_q(x;\theta)$, defined as the expectation value of the Pauli-Z observable on the last qubit, provides the variational flexibility.

The optimization of the quantum circuit parameters $\theta$ is performed using the Adam optimizer, a gradient-based optimization algorithm well-suited for noisy landscapes. A learning rate of $\eta = 0.005$ was employed, and the training was conducted over 500 iterations. The gradients are computed using the parameter-shift rule, which enables efficient and exact differentiation of quantum circuits. The second-order spatial derivative required by the Bratu residual was approximated using the central finite-difference method with a discretization step of $h = 10^{-3}$, and the cost function was computed as the integrated squared residual over a uniform grid of $N = 100$ interior points in the domain $x \in (0,1)$.

## 4. Results and discussion

### 4.1 Results for the 1D Bratu equation

Figure 1 presents a comparison between classical and quantum solutions to the one-dimensional Bratu problem across a sequence of values of the continuation parameter $\lambda$. The classical solution, shown as a solid line in each subplot, is obtained using pseudo arc-length continuation (see Appendix A) and serves as a reference solution along the nonlinear branch. Superimposed on the black solid line are the red dots, representing the quantum variational solutions evaluated at discrete spatial locations using a fixed quantum ansatz.

At low values of $\lambda$, the quantum solutions closely match the classical profile, as expected in the near-linear regime of the Bratu problem. This agreement confirms that the variational quantum circuit is capable of accurately representing smooth and symmetric solutions using relatively few parameters. As $\lambda$ increases, the classical solution becomes more nonlinear and exhibits a sharp central peak due to the exponential term in the Bratu equation. In these regimes, the red quantum points continue to track the classical curve with high fidelity, demonstrating the expressiveness of the ansatz and the effectiveness of the optimization process.

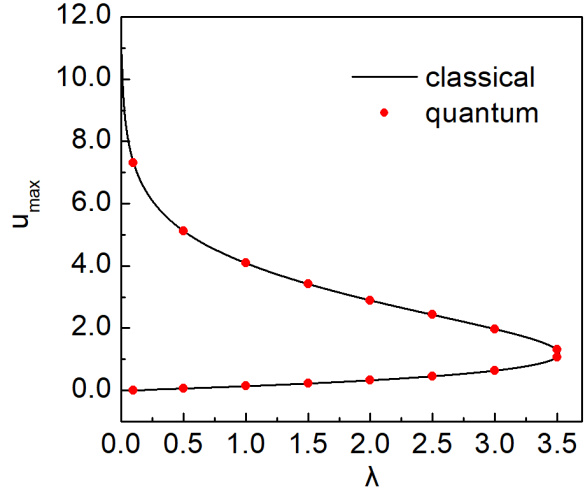

**Figure 1** Comparison of the classical and the quantum solver in different branches; $u_{max}(x = 0.5)$.

A key component enabling the quantum solver's performance is the design of the encoding function used to map classical data into the quantum state. In this study, each spatial coordinate is embedded into the quantum circuit via a differentiable encoding scheme. The choice of the encoding (see Eq. 3) directly impacts the circuit's ability to reconstruct steep features and reproduce the symmetry inherent in the Bratu solution. In particular, the encoding ensures that spatial structure is efficiently projected into the quantum Hilbert space, where it can be captured and reconstructed through parameterized rotations and entangling gates.

The red markers do not form a continuous curve but rather sample the quantum output at specific points along the spatial domain. Nevertheless, the alignment of these discrete values with the classical solution across all panels indicates that the quantum solver successfully approximates the spatial profile of $u(x)$ without participating in the continuation process itself. Instead, it operates as a variational corrector at fixed continuation parameters: given a fixed $\lambda$, the quantum circuit is trained to minimize the residual of the Bratu equation, yielding a discrete approximation $u_q(x; \theta)$ consistent with the PDE.

For the lower solution branch, the encoding function alone is sufficient to accurately recover the solution, and the $u_{pred}(x)$ term in Eq. 2 can be omitted. In contrast, capturing the upper branch requires incorporating the $u_{pred}(x)$ term as a first-order correction. This term guides the quantum model using information from the previous solution. The first upper-branch solution can be obtained via multi-start quantum optimization, after which the predictor–corrector procedure becomes fully quantum-driven. This strategy allows the quantum model to accurately follow the upper branch structure.

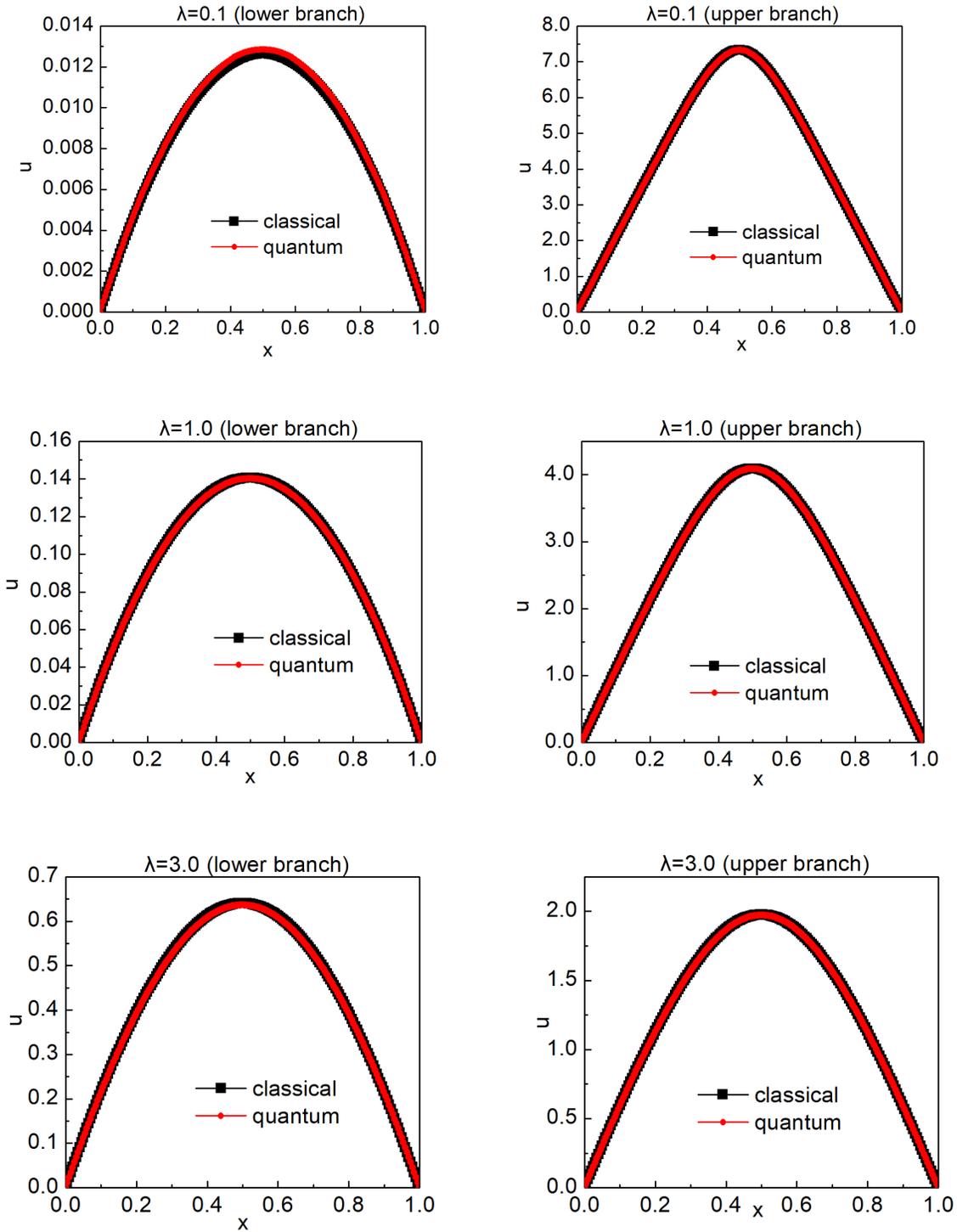

**Figure 2.** Comparison of the solution in the lower and upper branches, and near the critical (turning) point as computed via the VQA implementation.

Overall, the comparison shown in Fig. 1 demonstrates that quantum variational models can reliably reproduce classical solutions to nonlinear PDEs such as the Bratu equation — not only in the linear regime but also in highly nonlinear regions near turning points. The effectiveness of the encoding strategy plays a critical role in enabling generalization and resolution, supporting the broader application of quantum models to nonlinear scientific computing tasks.

For comparison reasons, we plot the results for $u$ along $x$ for $\lambda = 0.1$, 1.0 and 3.0 in the lower and upper branches for the classical and quantum solvers. The results are summarized in Fig. 2 and as it can be seen the results from coincide and are in very good agreement with previous works [25].

### 4.2 Quantum circuit optimization and weight behavior

The quantum circuit weights evolved differently depending on whether the solver was initialized toward the lower or upper branch of the Bratu solution. Below we compare the final optimized weights in both cases. The plots display the $R_X$, $R_Y$, and $R_Z$ rotation angles across all layers and qubits in the variational quantum circuit. The weight index on the x-axis refers to a flattened enumeration of all weight parameters in the circuit, combining layer number, qubit number, and rotation type. For example, index 0 corresponds to the $R_X$ weight of the first qubit in the first layer, index 1 corresponds to the $R_X$ weight of the second qubit in the first layer, and so on. This allows for direct visual comparison of all parameter values between the lower branch and upper branch.

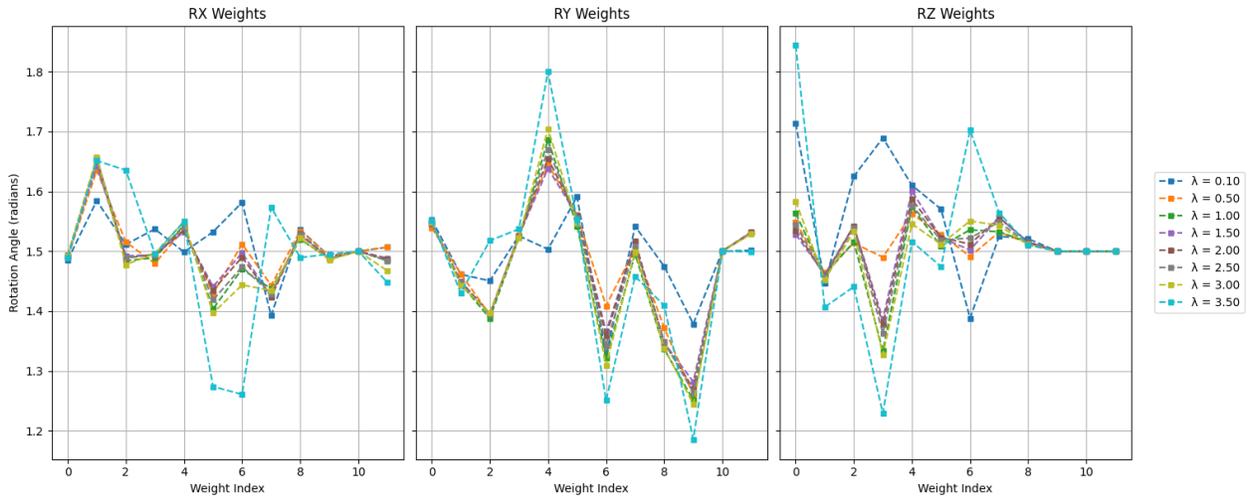

**Figure 3.** Optimized quantum circuit weights: Lower branch split by rotation type.

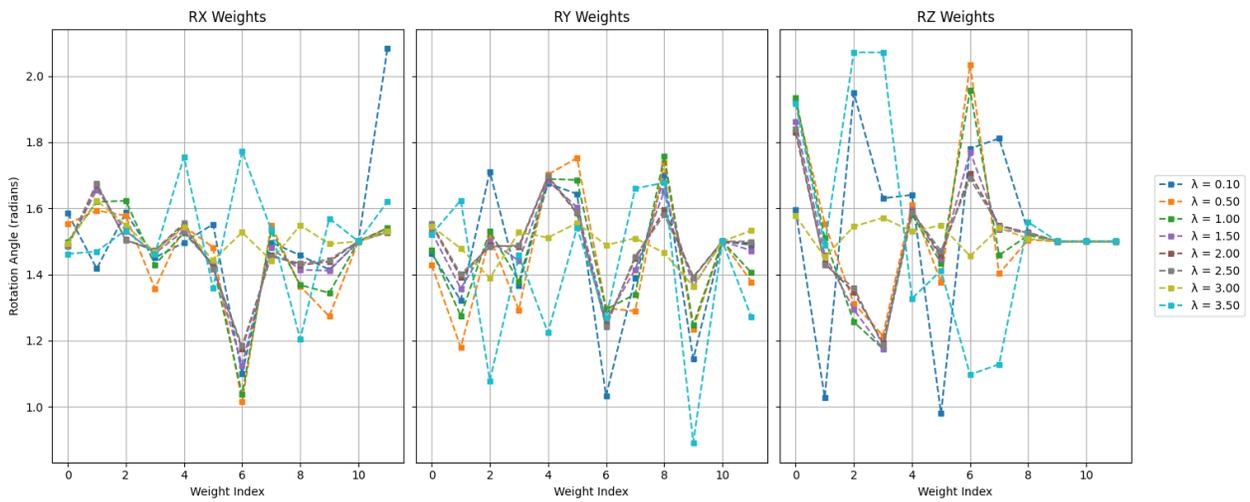

**Figure 4.** Optimized quantum circuit weights: Upper branch split by rotation type.

Across both the lower (Fig. 3) and upper branches (Fig. 4), the rotation angles display consistent patterns for small to moderate values of $\lambda$ (i.e., $\lambda \leq 2.00$). In this regime, the optimization process converges to relatively smooth and coherent parameter profiles, with minimal deviation among the different $\lambda$ values. This suggests that light regularization does not significantly constrain the expressivity of the circuit, and the optimizer finds similar parameter configurations regardless of the regularization strength.

However, as $\lambda$ increases beyond 2.00, especially for $\lambda = 3.00$ and $\lambda = 3.50$ notable deviations begin to emerge. These higher values of $\lambda$ promote stronger regularization, visibly flattening certain weights and amplifying the sparsity across the parameter space. In both branches, this leads to sharper dips and plateaus in the rotation angles, indicating the suppression of less critical rotations and possible pruning of redundant degrees of freedom.

While the general trends across $\lambda$ are shared, the two branches exhibit several key differences in their response to increasing regularization. In the lower branch, the rotation angles are generally more stable across weight indices, ranging approximately from 1.2 to 1.8 radians. The weight profiles for different $\lambda$ values largely overlap for most of the indices, with modest divergence at higher regularization levels. For $\lambda = 3.50$, the parameter trajectory displays more pronounced flattening toward the tail end (indices 28–35), but the variation across the earlier part of the circuit remains relatively limited.

In contrast, the upper branch demonstrates a broader dynamic range in its optimized parameters, with rotation angles spanning from as low as 0.9 to over 2.0 radians. The impact of $\lambda$ is more visually distinct here, with the optimized weights for larger $\lambda$\lambda$\lambda$ values diverging more sharply from those corresponding to weaker regularization. Specifically, $\lambda = 3.50$ exhibits both the most pronounced peaks and the deepest dips, suggesting that the upper branch undergoes more aggressive pruning and weight suppression under strong regularization. This could reflect a greater degree of parameter redundancy in the upper branch or a higher sensitivity of this subcircuit to the regularization term.

One striking observation is that the upper branch appears to tolerate—or even benefit from—stronger regularization without losing the ability to converge, whereas the lower branch seems more conservative, with fewer weights fully suppressed. This asymmetry may be related to the functional roles of each branch within the full circuit or differences in entanglement structure and data encoding responsibilities.

The analysis reveals that regularization acts as a tuning knob to sculpt the expressive capacity of the variational circuit. For small $\lambda$, both branches preserve most of their rotational degrees of freedom, potentially at the cost of overfitting or lack of generalization. As $\lambda$ increases, the circuit selectively suppresses weights—particularly in the upper branch—leading to simpler, more interpretable models. This supports the hypothesis that certain parts of the circuit can be pruned without compromising performance, potentially enhancing generalization and reducing resource overhead in near-term quantum hardware implementations. Moreover, the differential impact on the two branches highlights the value of architectural decomposition in variational circuits. Understanding which components of a quantum model are more amenable to sparsification provides a pathway toward efficient circuit design and tailored regularization strategies.

## 5. Conclusions and future work

This work demonstrates the successful application of a variational quantum algorithm (VQA) to solve the nonlinear one-dimensional Bratu equation. By embedding a parameterized quantum neural network (QNN) within a variational framework, we construct a quantum trial solution that satisfies the boundary conditions and transforms the differential equation into an optimization problem over quantum circuit parameters.

The method effectively captures both the lower and upper solution branches of the Bratu equation. In the lower branch regime, the quantum model accurately reproduces the classical solution without requiring prior information. In contrast, capturing the upper branch, which is characterized by sharp gradients and increased nonlinearity, benefits from a predictor–corrector strategy, in which guidance from the previous continuation step is used to steer the optimization toward the unstable solution branch. This strategy enables the quantum circuit to converge efficiently to high-amplitude solutions that would otherwise be difficult to reach through purely data-driven optimization.

Through extensive simulations on a noiseless quantum simulator, the results show strong agreement between the quantum and classical solvers across a range of $\lambda$ values, including near the critical turning point where solution behavior becomes highly nonlinear. The structure of the optimized quantum circuit weights reflects the increased complexity of the upper branch, revealing purposeful and interpretable adaptations of the ansatz to the underlying solution landscape.

Importantly, the implementation leverages only three qubits and a small number of encoding functions and circuit layers, indicating that expressive and accurate variational quantum solvers for nonlinear PDEs are achievable with minimal quantum resources. This highlights the potential for practical deployment of VQA-based methods on near-term quantum devices, particularly for structured scientific problems where classical solutions are costly or difficult to compute.

While the current results are promising, there are limitations. All experiments are conducted on a noiseless simulator, which does not account for decoherence, gate errors, or measurement noise that are prevalent in existing quantum hardware. The scalability of the method is also constrained by the limited number of qubits and circuit depth: increasing problem complexity or dimensionality will require deeper circuits and more expressive encodings, which may exacerbate noise sensitivity and training instability. Furthermore, the current optimization process relies on classical gradient descent and is sensitive to initialization—particularly for the upper solution branch. While classical guidance via low-fidelity solutions improves convergence, it also reduces the autonomy of the quantum solver. Incorporating more robust initialization schemes, adaptive encoding strategies, or quantum-aware regularization could help mitigate these issues.

Several promising directions emerge from this work. First, the method can be extended to higher-dimensional versions of the Bratu equation, which would involve generalizing the input encoding and ansatz architecture to multiple spatial variables. Second, future work will explore embedding pseudo arc-length continuation directly into the quantum optimization process by defining a parametric cost function over both $\theta$ and $\lambda$. Third, exploring hybrid noise-aware training strategies—such as error mitigation or circuit transpilation—will be essential for deploying this method on real quantum hardware. Additionally, the approach could be extended to other nonlinear PDEs, including reaction-diffusion systems, the Allen–Cahn equation, and Ginzburg–Landau models. Adapting the framework to support parametric PDE families or inverse problems (e.g., parameter inference) also presents a compelling research avenue, especially when coupled with probabilistic quantum models.

Overall, the study establishes a solid foundation for quantum-assisted solvers for nonlinear differential equations and suggests several future directions, including generalization to higher dimensions, more complex boundary conditions, and integration of continuation techniques directly into the quantum optimization process.

**APPENDIX A**

In classical computers, solving the 1D Bratu equation with the finite-difference method involves discretizing the differential equation while incorporating the boundary conditions. This approach converts the original problem, see Eq. (1), into a system of nonlinear equations, which are then solved using the Newton method. By employing a uniform grid with spacing $h$, where $h = \frac{L}{N+1}$, for an integer $N$, the domain $[0,1]$ is discretized as $x_i = ih$ for $i = 0,1,\ldots,N+1$. Utilizing a standard finite-difference scheme, the discretized formulation of the 1D Bratu equation reads,

$$\frac{u_{i+1} - 2u_i + u_{i-1}}{h^2} + \lambda e^{u_i} = 0 \tag{A1}$$

The nonlinear system created by the discretization of the set of equations via the finite differences, in a general case of a $N$-dimensional space, can be written as,

$$\mathbf{F}(\mathbf{u},\gamma) = 0 \tag{A2}$$

where $\mathbf{F}$ is the residual vector, $\mathbf{u}$ is the discretized form of the solution $u$. In the context of Newton-Raphson, the linearized system reads,

$$\frac{\partial \mathbf{F}}{\partial \mathbf{u}} \delta \mathbf{u} = -\mathbf{F} \tag{A3}$$

where $\frac{\partial \mathbf{F}}{\partial \mathbf{u}} = \mathbf{J}$, is the Jacobian and for the $i$-th equation reads,

$$J_{ij} = \begin{cases} -\frac{1}{h^2}, j = i-1 \text{ or } j = i+1 \\ -\frac{2}{h^2} + \gamma e^{u_i}, j = i \end{cases} \tag{A4}$$

The Bratu equation exhibits bifurcation behavior. As $\lambda$ increases, there comes a critical value $\lambda_c$, beyond which no solution exists. Below $\gamma_c$, there are two solution branches: A Lower branch witch stable solutions and an upper branch with unstable solutions. The turning point occurs at $\lambda_c$, where the solution curve (parameterized by $\lambda$) folds over. Standard methods, like Newton, fail near the turning point because the Jacobian matrix, Eq. (A3), becomes singular. To pass beyond the turning point and trace both solution branches, the pseudo-arc-length continuation algorithm [26,27] is introduced in the system as a new parameter. This method reparametrizes the problem using an auxiliary parameter $s$ (arc length) rather than directly using $\lambda$. This way, the solution $\boldsymbol{u}$ as well as the parameter $\gamma$ are expressed as functions of $s$, i.e. $(\boldsymbol{u}, \lambda) = [\boldsymbol{u}(s), \lambda(s)]$. This allows smooth traversal of the solution curve, even past the turning points. The residual vector, $\mathbf{F}$, is augmented by the arc-length equation, $\boldsymbol{\Phi}(\boldsymbol{u}, \lambda) = 0$, which implies the construction of an augmented Jacobian matrix, $\mathbf{J}_{aug}$,

$$\mathbf{J}_{aug} = \begin{bmatrix} \mathbf{J} & \dfrac{\partial \mathbf{F}}{\partial \lambda} \\ \dfrac{\partial \boldsymbol{\Phi}}{\partial \boldsymbol{u}} & \dfrac{\partial \boldsymbol{\Phi}}{\partial \lambda} \end{bmatrix} \tag{A5}$$